\documentclass[twocolumn,pra,superscriptaddress]{revtex4}
\usepackage{color}
\usepackage{amsmath}
\usepackage{amssymb}
\usepackage{graphicx}
\usepackage{esint}
\usepackage{mathrsfs}

\begin{document}

\preprint{APS/123-QED}

\title{Superposition of macroscopically squeezed states with enhanced squeezing in cavity optomechanical system at single-photon level}
\author{Sai-Nan Huai}
\affiliation{Institute of Microelectronics, Tsinghua University, Beijing 100084,
China}

\author{Wei Nie}
\affiliation{Institute of Microelectronics, Tsinghua University, Beijing 100084,
China}

\author{Yun-bo Zhang}
\affiliation{Institute of Theoretical Physics, Shanxi University, Taiyuan 030006,
China}

\author{Yu-xi Liu}
\email{yuxiliu@mail.tsinghua.edu.cn}
\affiliation{Institute of Microelectronics, Tsinghua University, Beijing 100084,
China}

\affiliation{Beijing National Research Center for Information Science and Technology (BNRist), Beijing 100084, China}

\date{\today}

\begin{abstract}

We propose an efficient approach to generate the superposed macroscopically squeezed states with enhanced squeezing in a two-mode optomechanical system. This can be achieved by introducing a sinusoidal modulation to either the cavity frequencies or the coupling strengths between two cavity modes. The squeezement of the oscillator can be significantly enhanced to 12.16 dB with single photon, once the relative ratio of coupling strength is optimized under proper conditions. Further enhanced squeezing can be obtained by carefully adjusting the system parameters. In terms of the Wigner quasi-probability distribution, we show the squeezed error ellipses and interference fringes of the Yurke-Stoler-type squeezed states, denoting the squeezing and superposition properties. Our state generation scheme show reliable performance and robust resistance to finite environmental fluctuations, which implies applications for both fundamental interest and practical value.

\end{abstract}

\maketitle


\section{INTRODUCTION} \label{Sec.III}

Quantum superposition~\cite{Romero2011,Asadian2014,Arndt2014,Pikovski2015}, which is the intrinsic distinction between the classical and quantum worlds, has attracted considerable interest since its discovery. In the past few years, various superpositions of quantum states, e.g., superposition of number or coherent states~\cite{Schrödinger1935, Leibfried2005, Ourjoumtsev2007, Duan2019, Johnson2017}, have been explored in various systems, ranging from micro-scale, such as laser-trapped ions~\cite{Monroe1996}, ultracold atoms~\cite{Berrada2013}, superconducting systems~\cite{Wal2000,Liu2005,Clarke2008,Devoret2013}, to meso-scale, e.g., microwave cavities interacting with Rydberg atoms~\cite{Noel1996,Brune1996, Raimond1997}. However, it remains freshly studied to generate macroscopically superposition states in macroscopic systems with up to $10^{10}$ atoms, e.g., qubit-oscillator~\cite{Liao2016PRA} and cavity optomechanical systems~\cite{Liao2016PRL, Hoff2016,Aspelmeyer2014}. Decoherence to such huge number of atoms, induced by the quantum and thermal fluctuations, can often destroy and make it challenging to preserve the superposition. Meanwhile, squeezed state appeals a lot to researchers due to its high performance in the aspects of the continuous variable quantum information processing~\cite{Caves1980,Lorenz2001} and the quantum metrology~\cite{Bocko1996,Bollinger1996,Kessler2014,Barish1999}. Its intrinsic property of quadrature squeezement permits the reduction of quantum fluctuation in one quadrature to below the ground state (vacuum) level. Particularly, it plays an important role in ultrasensitive measurement, such as the detection of gravitational wave~\cite{Aasi2013}, photon scattering recoil event at the single photon level~\cite{Hempel2013}, and possible future astronomical observations such as supernova explosions. The aspiration for higher measurement precision inspires researchers to explore further enhancement of squeezing to beyond the quantum limit. Taking overall consideration, the state engineering of the quantum superposition of squeezed states~\cite{Hach1993,Fan1994,Obada1997,Barbosa2000,El-Orany2011,Balamurugan2017,Sanders1998,Schleich1988,Xin1994} with both the macro-scale and enhanced squeezing degree, remains challenging but holds much promise for both the fundamental interest and practical value. 


Many pioneering schemes have been proposed during the process of improving the degree of squeezing. As shown in Ref.~\cite{LaHaye2004}, position resolution with a factor of 4.3 above the quantum limit is achieved at millikelvin temperatures. Efficient squeezing can also be achieved via the backaction evading measurement (BAE)~\cite{Clerk2008,Hertzberg2010}, or detuned mechanical parametric amplification (DMPA) with continuous weak measurement and feedback to the system~\cite{Szorkovszky2011}. Besides, the generation of squeezing can be realized via the reservoir engineering~\cite{Jahne2009,Kronwald2013,Kronwald2014,Woolley2014,Gu2013,Liao2018,Mari2009,Farace2012,Qu2014,Lei2016,Zhang2019}, which does not require any feedback. For example, when the cavity is driven by the squeezed light, a squeezed heat bath can be provided, and transferred to the membrane via the radiation pressure efficiently~\cite{Jahne2009}. An intermediate mode can also be engineered, via the modulated driving~\cite{Mari2009,Farace2012,Qu2014} induced beam-splitter-like interactions, and then serve as a reservoir to cool the Bogoliubov mode of the target system operators to the near ground state~\cite{Kronwald2013,Kronwald2014,Woolley2014,Gu2013,Liao2018,Zhang2019}. However, these schemes are somewhat constrained by the requirements of ultra-low temperature or absolute strong coupling to the cavity mode, which are kind of difficult to be realized in current experiments. Additional side-effect is the possible parametric instability, which may induce the system entering the chaotic regime~\cite{Monifi2016,Lu2015,Bakemeier2015}. Moreover, large squeezing can be achieved by increasing the effective coupling strength, in the way of achieving large number of average photons~\cite{Shi2013, Nunnenkamp2010}, or adding atoms into the cavity~\cite{Han2013}, i.e., the atom-assisted squeezing. Nevertheless, the growing number of photons or atoms is accompanied with intensive dissipation effects, actually leading to the increase of threshold temperature of the mirror's bath. Thus, in this paper, we focus on the single-photon induced squeezing to avoid from the intensive dissipation caused by large number of photons. Meanwhile, we resort to the modulated scheme~\cite{Liao2014,Liao2015,Liao2016PRA,Liao2016PRL,Hoff2016,Aspelmeyer2014,Huang2017,Huang2019,Wu2018,Quijandria2018}, i.e., modulating the system parameters, such as frequencies or coupling strengths, to optimize relative ratios of coupling strengths, rather than simply increase their magnitudes.

Motivated by works in Refs.~\cite{Liao2014,Liao2015,Liao2016PRA,Liao2016PRL,Hoff2016,Aspelmeyer2014,Huang2017,Huang2019,Wu2018,Quijandria2018}, here we revisit the standard quadratically coupled optomechanical system~\cite{Murch2008,Purdy2010,Thompson2008} with the so-called ``membrane-in-the-middle'' (MIM) configuration~\cite{Thompson2008,Teng2012,Bai2016}, and propose an efficient approach to generate the superposed macroscopically squeezed states with enhanced squeezing. This goal can be realized by applying a sinusoidal modulation to the cavity frequency or the photon hopping rate and cavity optomechanical coupling strengths. Once the modulating frequencies and amplitudes satisfy specific conditions, the system can be approximately described by a quadratic-coupling optomechanical system with modified coupling strengths and mechanical resonator frequency. Then an effective strong coupling regime is achieved and the squeezement of the mechanical resonator is enhanced remarkably below the zero-point level to 12.16 dB. As far as we know, this degree of squeezing goes beyond to the best experimental realization 11.5 dB in Refs.~\cite{Vinante2013,Agarwal2016}. By carefully designing the initial state of the system and measuring the cavity at proper time, we can arrive at the superposed squeezed state of mechanical mode. We will show its enhanced squeezing and superposition properties in terms of Wigner function. We will also show the robust performance of this state generation scheme in open system.

The rest of our paper is organized as follows. In Sec.~\ref{Sec.system}, we first give an explicit description for the quadratically coupled MIM optomechanical system. Then, the effective Hamiltonian in presence of a sinusoidal modulation to cavity frequencies are presented in Sec.~\ref{Mfrequency}. In Sec.~\ref{enhance}, we show the mechanism of enhanced squeezing with single photon, and verify its validity through fidelity. In Sec.~\ref{superpose}, we study the generation of superposed enhanced macroscopically squeezed states and its interference effect in terms of Winger quasi-probability distribution. In Sec.~\ref{dissipation}, we consider the effect of environmental fluctuations to the performance of the state generation program. Finally, we summarize our research results in Sec.~\ref{conclusion}. Besides, the explicit processes with modulated photon hopping rate and optomechanical strengths are shown explicitly in Appendix~\ref{Mcoupling}. Meanwhile, the details in calculating the Wigner distributions are shown in Appendix~\ref{wigner}.

\section{THEORETICAL MODEL}\label{Sec.system}
\begin{figure}[ptb]
\includegraphics[scale=0.25,clip]{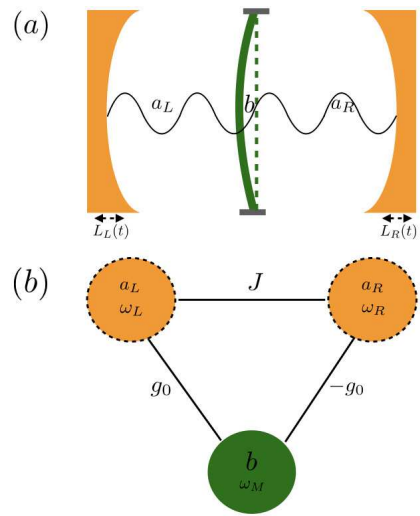}\
 \caption{(Color online) (a) Schematic diagram of a modulated hybrid membrane-in-the-middle (MIM) configuration of cavity optomechanical system. A thin membrane, with the phonon mode $b$ and located inside the cavity, separates the cavity into left and right modes, i.e., $a_L$, $a_R$. The cavity frequencies are modulated by the adjustion of cavity lengths $L_L(t)$ and $L_R(t)$. (b) Equivalent coupled-harmonic-resonator model. The parameters $\omega_L$, $\omega_R$, and $\omega_M$ represent the resonant frequencies of left, right cavities and mechanical resonator, respectively. The strength $J$ denotes the photon tunneling between the two cavities, while $g_0$ is the quadratic cavity optomechanical coupling.}%
\label{Figure1}%
\end{figure}
The standard quadratically coupled MIM optomechanical system we considered is schematically shown in Fig.~\ref{Figure1}(a), where a thin dielectric membrane is placed inside a Fabry-P\'{e}rot cavity. We assume that the reflection coefficient of the membrane is large enough to separate the cavity into the right and left halves, and permits photon tunneling with a coefficient $J$ at the same time. The independent modes in these two halve cavities are labeled by $R$ and $L$. We also assume that the membrane is located in the node of the intra-cavity standing wave, such that the cavity field is coupled quadratically to the mechanical displacement~\cite{Aspelmeyer2014}. The equivalent coupled-harmonic-resonator model is shown in Fig.~\ref{Figure1}(b), i.e., the right, left cavities and the mechanical resonator mutually interact. The Hamiltonian of this hybrid MIM optomechanical system can be described as 
\begin{eqnarray}
H_0&=&\omega _{c} (a_{L}^{\dagger }a_{L}+a_{R}^{\dagger
}a_{R})+J\left( a_{L}^{\dagger }a_{R}+a_{R}^{\dagger }a_{L}\right)  \nonumber \\
&&+\omega_{M}b^{\dagger }b+g_{0}\left( a_{L}^{\dagger }a_{L}-a_{R}^{\dagger
}a_{R}\right) \left( b^{\dagger }+b\right) ^{2}. \label{H0}
\end{eqnarray} 
Here we have assumed the same frequency $\omega_c$ for the left and right cavity modes $a_{L}$ and $a_{R}$, and $\omega_M$ is the resonance frequency for mechanical mode $b$. Specifically, the quadratic cavity-membrane coupling strength $g_0={\tau\zeta^2}[(1-T)/{T}]^{1/2}/2$~\cite{Teng2012,Bai2016}, with $T$ denoting the membrane transmissivity. The parameter $\tau=2L/c$ signifies the round trip time for each cavity, and $\zeta=\omega_c/L$, with $L$ the cavity length when the membrane and the outside mirror remain static. In fact, our model is general and system independent, it can be implemented with various experimental setups~\cite{Murch2008,Purdy2010,Thompson2008,Xue2007}. In addition to the MIM configuration~\cite{Thompson2008}, the cold-atom optomechanical system~\cite{Murch2008,Purdy2010} and superconducting quantum interference device (SQUID)~\cite{Xue2007} are also potential platforms to demonstrate our scheme with carefully design.

In order to get the effective strong coupling and enhanced squeezing, we introduce a sinusoidal modulation to the cavity frequencies $\omega_c$ or the photon-hopping interaction constant $J$ and optomechanical coupling constant $g_0$. We note that this method has been widely applied to various tasks, such as generating distinct coherent states~\cite{Liao2016PRA,Liao2016PRL}, manipulating counter-rotating interactions~\cite{Huang2017} to achieve ultrastrong Jaynes-Cummings interaction~\cite{Huang2019}, realizing efficient and compact switch for quantum circuits~\cite{Wu2018}, and designing $\mathcal{PT}$-symmetric circuit QED~\cite{Quijandria2018}. Here, we will mainly study the situation where a sinusoidal modulation is introduced to the cavity frequencies, while the case when the modulations are apllied to the photon-hopping rate $J$ and optomechanical coupling $g_0$ will be shown in Appendix~\ref{Mcoupling} explicitly. 

\section{HAMILTONIAN ENGINEERING}\label{Mfrequency}
Now we assume that the cavity frequencies of the left and right cavities are modulated as 
\begin{eqnarray}
\omega _{L} (t)&=&\omega _{c}+\Delta _{0}\cos 2Jt, \\
\omega _{R} (t)&=&\omega _{c}-\Delta _{0}\cos 2Jt,
\end{eqnarray}%
respectively, and the modulated Hamiltonian is given by
\begin{equation}
H_M(t)=\Delta _{0}\cos 2Jt\left(a_{L}^{\dagger }a_{L}-a_{R}^{\dagger }a_{R}\right).
\end{equation}
Here, we assume that the modulating amplitude $\Delta_0$ is far less than the modulation frequency $J$, i.e., $\Delta_0\ll J$. Considering that cavity frequency is proportional to $L^{-1}$~\cite{Aspelmeyer2014}, the modulated cavity frequencies can be realized by modulating the lengths of the left and right cavities as $L\left[1-\left(\Delta _{0}\cos 2Jt\right)/2\omega_c\right]^2$ and  $L\left[1+\left(\Delta _{0}\cos 2Jt\right)/2\omega_c\right]^2$, respectively. In terms of experimental practicability, the cavity lengths can be modulated with the help of two lasers with beat frequency $2J$~\cite{Schmidt2015}, or the piezoelectric transducer, which converts the electrical signal to mechanical vibration~\cite{Svelto1982}. Besides, the rapid response of cavity frequencies makes it reasonable to neglect the back-action of the length modulation to the membrane. Thus, the full Hamiltonian of this hybrid MIM system under the frequency modulation can be engineered as 
\begin{eqnarray}
H(t)&=&H_0+H_M(t)\nonumber\\
&=&\omega_{L}(t) a_{L}^{\dagger }a_{L}+\omega _{R}(t) a_{R}^{\dagger
}a_{R}+J\left( a_{L}^{\dagger }a_{R}+a_{R}^{\dagger }a_{L}\right)  \nonumber \\
&&+\omega_{M}b^{\dagger }b+g_{0}\left( a_{L}^{\dagger }a_{L}-a_{R}^{\dagger
}a_{R}\right) \left( b^{\dagger }+b\right) ^{2}. \label{H}
\end{eqnarray} 

The state-evolution of this system can be described by the Schr\"{o}dinger equation
\begin{equation}
i\frac{\partial }{\partial t}\left\vert \psi \left( t\right) \right\rangle
=H\left( t\right) \left\vert \psi \left( t\right) \right\rangle. \label{psi}
\end{equation} 
Usually, this equation can be solved by introducing an unitary transformation $Q(t)$, i.e., $\left\vert \psi \left( t\right) \right\rangle=Q(t)\left\vert \tilde{\psi} \left( t\right) \right\rangle.$ As a consequence, Eq.~(\ref{psi}) can be rewritten as 
\begin{equation}
i\frac{\partial }{\partial t}\left\vert \tilde{\psi} \left( t\right) \right\rangle
=\tilde{H}\left( t\right) \left\vert \tilde{\psi} \left( t\right) \right\rangle,
\end{equation} 
with $\tilde{H}(t)$ 
determined by~\cite{Scully1997} 
\begin{equation}
\tilde{H}(t)=Q^{\dagger}(t)H(t)Q(t)-iQ^{\dagger}(t)\frac{\partial Q(t)}{\partial t}.\label{Htilde}
\end{equation} 
In our case, the unitary transformation $Q(t)$ is defined as $Q(t)=P_1(t)P_2(t)P_3(t)$~\cite{Silveri2017}, with
\begin{eqnarray}
P_{1}\left( t\right) &=&e^{-i\omega _{c}\left( a_{L}^{\dagger}a_{L}+a_{R}^{\dagger }a_{R}\right) t}, \label{P1}\\ 
P_{2}\left( t\right) &=&e^{-iJ\left( a_{L}^{\dagger }a_{R}+a_{R}^{\dagger}a_{L}\right)t}, \label{P2}\\
P_{3}\left( t\right) &=&e^{-i\left[ \frac{\Delta _{0}}{2}\left(a_{L}^{\dagger }a_{L}-a_{R}^{\dagger }a_{R}\right) +\omega _{M}b^{\dagger }b%
\right] t}.\label{P3}
\end{eqnarray}
With the transformation $Q(t)$, the Hamiltonian shown in Eq.~(\ref{Htilde}) can be written explicitly as 
\begin{eqnarray}
\tilde{H}(t)&&=\left( a_{L}^{\dagger
}a_{L}-a_{R}^{\dagger }a_{R}\right)\left[\frac{\Delta _{0}}{4}F_1(b)+gF_2(b)+\rm{H.c.}\right]\nonumber\\
&&+\left(a_{L}^{\dagger }a_{R}e^{i\Delta_0t}-\rm{H.c.}\right)\left[-\frac{\Delta _{0}}{4}F_1(b)+gF_3(b)-\rm{H.c.}\right],\nonumber\\ \label{HTILDE}
\end{eqnarray}
with the functions 
\begin{eqnarray}
F_1(b)&=&e^{4iJt},\nonumber\\
F_2(b)&=&b^{2}\left(e^{-2i\delta_{-} t}+e^{-2i\delta_{+} t}\right)+\left( 2b^{\dagger }b+1\right) e^{-2iJt},\nonumber\\
F_3(b)&=&b^{2}\left(e^{-2i\delta_{+} t}-e^{-2i\delta_{-} t}\right)+\left( 2b^{\dagger }b+1\right)e^{-2iJt}.\nonumber
\end{eqnarray}

Besides, we have defined the modified coupling constant $g$ and detuning $\delta_{\pm}$ as,
\begin{equation}
g=\frac{g_0}{2},\ \ \  \delta_{\pm}=\omega_M\pm J,
\end{equation}
respectively. It is obvious that more interaction components show up due to cavity frequencies modulation, e.g., terms whose frequencies or amplitudes are related to $\Delta_0$ and $J$. However, we can always find a way to manipulate the desired terms into near-resonant ones, such that the resulted effective strong coupling terms actually dominate the evolution of the system, while other fast-osillating and relatively weak coupling ones can be neglected. In our work here, we would like to study the generation of the squeezing states for mechanical resonator, thus we consider the following conditions 
\begin{eqnarray}
J\gg\frac{5}{16}\Delta_0,  \ \ \    \Delta_0&\gg& 2g, \left\vert\delta_{-}\right\vert \nonumber\\
 \left\vert\Delta_0-2J\right\vert&\gg& g.  \label{condition}
\end{eqnarray}
In this situation, the terms other than those with frequency $\delta_{-}$ in Eq.~(\ref{HTILDE}) turn out to be fast-oscillating and weak coupling, while the coupling coefficients in the desired near-resonant terms are comparable to the oscillating frequencies, i.e., the effective strong coupling is realized. Thus, it is reasonable to conduct the rotating-wave approximation (RWA) to Eq.~(\ref{HTILDE}), leading to 
\begin{equation}
\tilde{H}_{\rm{eff}}(t)\simeq g\left( a_{L}^{\dagger
}a_{L}-a_{R}^{\dagger }a_{R}\right) \left( b^{2}e^{-2i\delta t}+b^{\dagger
2}e^{2i\delta t}\right).  \label{Heff'}
\end{equation}
For convenience, we have neglected the subscript and rewritten $\delta_{-}$ as $\delta$ to represent the slow oscillating frequency, i.e., $\delta=\delta_{-}=\omega_M-J$. We can further arrive at the effective Hamiltonian 
\begin{eqnarray}
H_{\rm{eff}}\left( t\right)&=&\omega _{c}\left( a_{L}^{\dagger }a_{L}+a_{R}^{\dagger }a_{R}\right)
+\delta b^{\dagger }b  \nonumber\\
&&+g\left( a_{L}^{\dagger }a_{L}-a_{R}^{\dagger
}a_{R}\right) \left( b^{2}+b^{\dagger 2}\right). \label{Heff}
\end{eqnarray}
by performing the unitary transformation $P_4(t)$, which is defined as 
\begin{equation}
P_{4}\left( t\right) =e^{i\left[ \omega _{c}\left( a_{L}^{\dagger
}a_{L}+a_{R}^{\dagger }a_{R}\right) +\delta b^{\dagger }b\right] t}. \label{P4}
\end{equation}

As a result, under the conditions shown in Eq.~(\ref{condition}), we engineer an effective strong quadratically coupled Hamiltonian, i.e., Eq.~(\ref{Heff}), for the modulated cavity frequency case. Additionally, the conditions and engineered Hamiltonian correspond to the modulated coupling strengths case are shown in Eqs.~(\ref{condition2}) and~(\ref{Heff2}), respectively. 
The flexibility of modulation permits the detuning $\delta=\omega_M-J$ to be small enough than the modified coupling strength $g$, which further leads to effective strong and even ultra-strong quadratically coupling. This Hamiltonian engineering mechanism actually lays the foundation of generating the superposed macroscopically squeezed states with enhanced squeezing at single-photon level, which we would like to give a detailed description in the rest part of our work.
\section{ENHANCED SQUEEZING WITH SINGLE-PHOTON}\label{enhance}
Let us first focus on how to enhance the degree of squeezing for the states of the mechanical resonator utilizing the Hamiltonian in Eq.~(\ref{Heff}). We hope that the enhanced squeezing can go far beyond the quantum limit than most of reported schemes. 
With carefully designed initial state and dynamical evolution for a proper time, we can obtain the enhanced squeezing state for the mechanical mode. 

For a closed system without dissipation, we can get analytical results, according to the effective Hamiltonian $H_{\rm{eff}}\left( t\right)$ in Eq.~(\ref{Heff}) and the corresponding time-dependent Schr\"{o}dinger equation $i{\partial }\left\vert \psi_{\rm{eff}}\left( t\right) \right\rangle/{\partial t}
=H_{\rm{eff}}\left( t\right) \left\vert \psi_{\rm{eff}}\left( t\right) \right\rangle$. Here the instantaneous state $\left\vert\psi_{\rm{eff}}\left( t\right)\right\rangle$ is related to the analytical state $\left\vert\psi\left( t\right)\right\rangle$ in Eq.~(\ref{psi}) with several unitary transformations, which have been shown in Eqs.~(\ref{P1})-(\ref{P3}) and~(\ref{P4}), i.e., 
\begin{equation}
\left\vert\psi\left( t\right)\right\rangle\simeq P_1\left( t\right)P_2\left( t\right)P_3\left( t\right)P_4\left( t\right) \left\vert\psi_{\rm{eff}}\left( t\right)\right\rangle. \label{psit}
\end{equation}  

Using the standard disentangling techniques utilized in quantum optics~\cite{Mufti1993}, we can evaluate the time evolution operator, corresponding to the effective Hamiltonian $H_{\rm{eff}}$, as
\begin{equation}
e^{-iH_{\rm{eff}}t}=e^{-i\omega _{c}\left( a_{L}^{\dagger }a_{L}+a_{R}^{\dagger
}a_{R}\right) t}e^{\frac{i}{2}\delta t}e^{\frac{1}{2}\left( \xi ^{\ast
}b^{2}-\xi b^{\dagger 2}\right) }e^{i\eta \left( b^{\dagger }b+\frac{1}{2}%
\right)}. \label{Hefft}
\end{equation}
The time-dependent parameters $\xi\equiv\xi \left( t\right)$ and phase factor $\eta\equiv\eta \left( t\right)$ are defined as
\begin{eqnarray}
\xi \left( t\right)&=&e^{i\left( \eta +\frac{\pi }{2}\right) }\sinh
^{-1}\left( \frac{r\sin \chi t}{\chi }\right), \label{xi}\\
\eta \left( t\right)&=&\tan ^{-1}\left( \frac{p\tan \chi t}{2\chi }\right),\label{eta}
\end{eqnarray}
respectively. It is obvious that $\xi \left( t\right)$ represent the degree of the squeezing as shown in quantum optics~\cite{Mufti1993}. We also note that $\eta\left(t\right)$ is a real number, and the other parameters can be evaluated as
\begin{eqnarray}
p &=&-2\delta, \\
r({N}_{LR})&=&2g{N}_{LR}, \label{rNLR}\\
\chi({N}_{LR}) &=&\sqrt{\delta ^{2}-4g^{2}{N}_{LR} ^{2}}.\label{chiNLR}
\end{eqnarray}
Here, ${N}_{LR}=a_{L}^{\dagger }a_{L}-a_{R}^{\dagger }a_{R}$ denotes the photon number inversion between the left and right cavities. The Eqs.~(\ref{xi})-(\ref{chiNLR}) indicate that the degree of squeezing is related to photon number. For the convenience of calculation and in oder to reduce the intensive dissipative effects caused by large number of photons for the open system which we will study in Sec.~\ref{dissipation}, we will mainly consider the squeezing of the mechanical resonator with single photon, i.e., $n_L+n_R=1$. Here, $n_L~(n_R)=0,1,2\cdots$ denotes the photon excitation number corresponding to the photon number operator $a_{L}^{\dagger }a_{L}~(a_{R}^{\dagger }a_{R})$ in the left (right) cavity. According to Eqs.~(\ref{psit}) and~(\ref{Hefft}), the dynamical change from the initial state $\left\vert\psi\left( 0\right)\right\rangle$ to the final state $\left\vert\psi\left( t\right)\right\rangle$, i.e., $\left\vert\psi\left( 0\right)\right\rangle\to\left\vert\psi\left( t\right)\right\rangle$, can be described as 
\begin{equation}
\left\vert \psi \left( t\right) \right\rangle\simeq U(t)\left\vert \psi \left( 0\right)\right\rangle.       \label{psitTheory}
\end{equation} 
Here, 
\begin{equation}
U(t)=P_1(t)P_2(t)P_3(t)P_4(t)e^{-iH_{\rm{eff}}t} \label{Ut}
\end{equation}
denotes the approximate time evolution operator associated with the original total Hamiltonian in Eq.~(\ref{H}). Meanwhile, the relation $\left\vert\psi\left( 0\right)\right\rangle\equiv\left\vert\psi_{\rm{eff}}\left( 0\right)\right\rangle$ has been used. 

In the process of getting the approximate analytical state $\left\vert \psi \left( t\right) \right\rangle$ in Eq.~(\ref{psitTheory}), rotating-wave approximation has been made under the condition Eq.~(\ref{condition}). Thus we need to verify its validity by checking the fidelity between the approximate analytical state $\left\vert \psi \left( t\right) \right\rangle$, and the numerical state $\left\vert \Psi \left( t\right) \right\rangle$ associated with the original total Hamiltonian in Eq.~(\ref{H}). The numerical results can be obtained by solving the Schr\"{o}dinger equation $i{\partial }\left\vert \Psi \left( t\right) \right\rangle/{\partial t}
=H\left( t\right) \left\vert \Psi \left( t\right) \right\rangle $ based on the general single-photon state of the system 
\begin{equation}
\left\vert \Psi \left( t\right) \right\rangle =\sum\limits_{n=0}^{\infty } 
\left[ A_{m}\left( t\right) \left\vert 1\right\rangle _{L}\left\vert
0\right\rangle _{R}+B_{m}\left( t\right) \left\vert 0\right\rangle
_{L}\left\vert 1\right\rangle _{R}\right] \left\vert m\right\rangle _{M}.  \label{Psit}
\end{equation}
Here $A_m(t)$ and $B_m(t)$ denote the probability amplitudes of the states $\left\vert 1\right\rangle _{L}\left\vert0\right\rangle _{R}\left\vert m\right\rangle _{M}$ and $\left\vert 0\right\rangle _{L}\left\vert1\right\rangle _{R}\left\vert m\right\rangle _{M}$, respectively. The subscripts $L$, $R$, and $M$ represent the left, right cavities, and the mechanical resonator, respectively. The equations of motion can be expressed as  
\begin{eqnarray}
\dot{A}_{m}\left( t\right) &=&-i\left( \omega _{c}+\Delta _{0}\cos
2Jt+m\omega _{M}\right) A_{m}\left( t\right) -iJB_{m}\left( t\right)  \nonumber\\
&&-ig_{0} [ \left( 2m+1\right) A_{m}\left( t\right) +\sqrt{m\left(
m-1\right) }A_{m-2}\left( t\right) \nonumber\\
&&+\sqrt{\left( m+1\right) \left( 
m+2\right) }A_{m+2}\left( t\right) ],\label{Amt} \\  
\dot{B}_{m}\left( t\right) &=&-i\left( \omega _{c}-\Delta _{0}\cos
2Jt+m\omega _{M}\right) B_{m}\left( t\right) -iJA_{m}\left( t\right)  \nonumber\\
&&+ig_{0} [ \left( 2m+1\right) B_{m}\left( t\right) +\sqrt{m\left(
m-1\right) }B_{m-2}\left( t\right) \nonumber\\
&&+\sqrt{\left( m+1\right) \left(
m+2\right) }B_{m+2}\left( t\right) ]. \label{Bmt}
\end{eqnarray}%
As long as the coefficients $A_m(0)$ and $B_m(0)$ $(m\geq 0)$ for the initial state $\left\vert \Psi \left( 0\right) \right\rangle$ are given, we can obtain the state $\left\vert \Psi \left( t\right) \right\rangle $ at any given time by solving Eqs.~(\ref{Amt}) and~(\ref{Bmt}) numerically.
\begin{figure}[ptb]
\includegraphics[scale=0.39,clip]{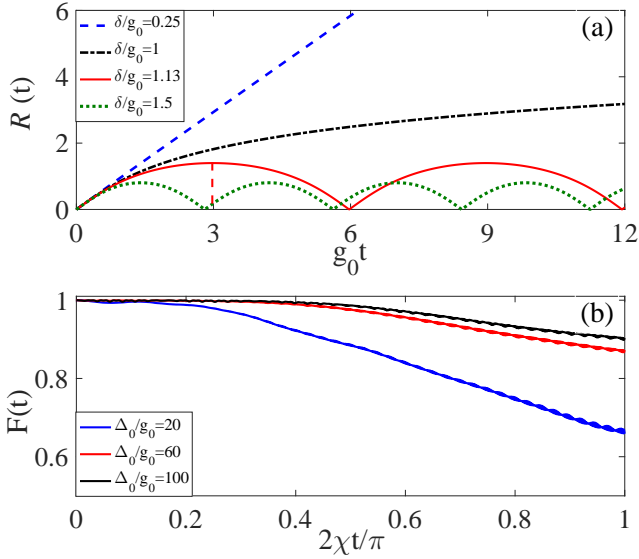}\
 \caption{(Color online) (a) The degree of mechanical squeezing $R(t)$ under different rescaled detunings, e.g., $\delta/g_0=0.25,\,1,\,1.13,\,1.5$. Here, the rescaled modulating amplitude $\Delta_0/g_0=100$. (b) The fidelity $F(t)$ for the given detuning $\delta/g_0=1.13$, under different rescaled modulating amplitude $\Delta_0/g_0=20,\,60,\,100$. The rescaled photon tunneling rate $J/g_0=398.6$.}%
\label{Figure2}%
\end{figure}
We now consider the mechanical squeezing with single-photon by assuming the initial state $\left\vert \psi \left( 0\right) \right\rangle=\left\vert \Psi \left( 0\right) \right\rangle=\left\vert 1\right\rangle_L\left\vert 0\right\rangle_R\left\vert 0\right\rangle_M$, i.e., there is only one photon in the left cavity while vacuum in the right one, and the mechanical membrane is in the ground state $\left\vert 0\right\rangle_M$. According to Eq.~(\ref{psitTheory}), the wave function $\left\vert \psi(t)\right\rangle$ at time $t$ can be analtically written as 
\begin{eqnarray}
\left\vert \psi \left( t\right) \right\rangle&=&e^{-i\omega _{c}t}e^{-i\theta \left( t\right) }[ \cos \left(
Jt\right) \left\vert 1\right\rangle _{L}\left\vert 0\right\rangle _{R} \nonumber \\
&&-i\sin\left( Jt\right) \left\vert 0\right\rangle _{L}\left\vert 1\right\rangle _{R}%
] \left\vert Z\left( t\right) \right\rangle _{M},
\end{eqnarray}
with the global phase factor
\begin{equation}
\theta \left( t\right)=-\left[ \delta t+\eta \left( t\right)-\Delta _{0}t\right]/2. \label{thetat}
\end{equation}
Here $ \left\vert Z\left( t\right) \right\rangle _{M}$ is the squeezed vacuum state of the membrane with 
\begin{equation}
Z\left( t\right)=\xi \left( t\right) e^{-2i\left( \omega _{M}-\delta
\right) t}, 
\end{equation}
and the parameter $\xi \left( t\right)$ have been given in Eq.~(\ref{xi}). Meanwhile, the corresponding parameters in Eqs.~(\ref{rNLR}) and~(\ref{chiNLR}) can be specifically written as
\begin{eqnarray}
r=2g,\text{ \ \ \ }\chi =\sqrt{\delta ^{2}-4g^{2}},
\end{eqnarray} 
based on the single photon assumption shown in Eq.~(\ref{Psit}).
An expression of the squeezed vacuum state $\left\vert Z\left( t\right)\right\rangle_M$ in terms of the number state $\left\vert m\right\rangle$ is given by~\cite{Zayed2005,Gerry2004}
\begin{equation}
\left\vert Z\left( t\right)\right\rangle_M=\sum\limits_{m=0}^{\infty }S_{m}\left( Z\left( t\right)\right) \left\vert2m\right\rangle,
\end{equation}%
where
 \begin{equation}
S_{m}\left( Z\left( t\right)\right)=\frac{\left( -1\right) ^{m}}{\sqrt{\cosh R\left( t\right)}}\frac{\sqrt{\left( 2m\right) !}}{2^{m}m!}e^{im\Phi\left( t\right) }\tanh ^{m}R\left( t\right).
\end{equation}
\begin{figure}[ptb]
\includegraphics[scale=0.39,clip]{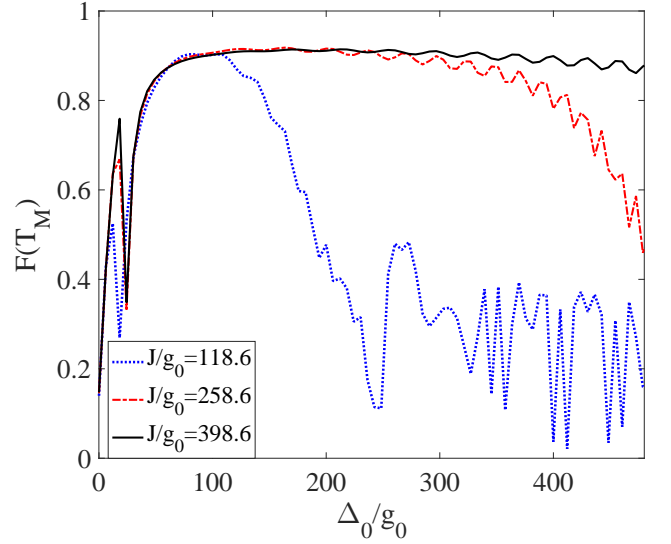}\
 \caption{(Color online) The fidelity $F(T_M)$ at the time $T_M$, where the maximum squeezing happens, versus the rescaled modulating amplitude $\Delta_0/g_0$, under different rescaled photon tunneling rates $J/g_0=118.6,\, 258.6, \,398.6$. Here the rescaled detuning $\delta/g_0$ is settled down to 1.13. The mechanical frequency is determined by $\omega_M=\delta+J$ for different tunnel rates $J$.}%
\label{Figure3}%
\end{figure}
Here the modulus ${R}\left( t\right)$ and argument $\Phi\left( t\right)$ of $Z\left( t\right)$ are defined as
\begin{eqnarray}
{R}\left( t\right)&=&\left\vert \sinh^{-1}\left( \frac{r\sin \chi t}{\chi }\right)\right\vert,\\
\Phi \left( t\right)&=&\arg \left[ Z\left( t\right)\right],
\end{eqnarray}
respectively. It is obvious that if $\delta> 2g$, then $\chi>0$, and ${R}\left( t\right)$ shows up as a periodical function, with the period $T=\pi/\chi$. The first maximum squeezing ${R}\left( T_M\right)=\sinh^{-1}\left( {r}/{\chi }\right)$ is achieved at $T_M=T/2$, as pointed out by the vertical red-dotted line in Fig.~\ref{Figure2}(a). Besides, we also show how the degree of squeezing ${R}\left( t\right)$ varies versus time, under different detunings $\delta$ in Fig.~\ref{Figure2}(a). It is shown that with the decreasing of the $\delta$, e.g., from $\delta/g_0=1.5$ (green dotted line) to 1.13 (red solid line), the period $T$ gets longer and ${R}\left( T_M\right)$ becomes larger. At the singular point $\delta=2g$, ${R}\left( t\right)=\sinh^{-1}\left(rt\right)$ grow exactly in the form of hyperbolic sine function (black dash-dotted line). As for the smaller detuning case $\delta< 2g$, for example, $\delta/g_0=0.25$ (blue dashed line), the mechanical squeezing ${R}\left( t\right)= \sinh^{-1}\left( {r\sinh \left({\sqrt{4g^{2}-\delta ^{2} } t}\right)}/{\sqrt{4g^{2}-\delta ^{2} }}\right)$ grows monofonically with time until the Hamitonian breaks down. We find that, although the larger detuning case ($\delta>2g$), which corresponds to periodic squeezing, can not get the given squeezing in shortest time, it shows higher stability to the perturbation around $T_M$. Thus, in this work, without loss of generality, we mainly focus on the larger detuning case with periodic squeezing, e.g., $\delta/g_0=1.13$ (red solid line). In this case, the maximum degree of mechanical squeezing can be enhanced to $-10{\rm{log}}_{10}[{\rm{exp}}(-2*R(T_M))]\simeq12.16$ dB~\cite{Scully1997}. 
Next, we will show its reliability by checking the fidelity between the analytical and numerical states.

According to Eq.~(\ref{Psit}), the initial state $\left\vert \Psi \left( 0\right) \right\rangle$ can be expressed as
\begin{equation}
A_0(0)=1,\text{\ \ \ } A_{m>0}(0)=B_{m\geq0}=0. 
\end{equation}
Then by solving the equations of motion in Eqs.~(\ref{Amt}) and~(\ref{Bmt}) numerically, we can get the state at any given time $\left\vert \Psi \left( t\right) \right\rangle$. In terms of the analytical expression $\left\vert \psi \left( t\right) \right\rangle$ and numerical result $\left\vert \Psi \left( t\right) \right\rangle$, the fidelity $F\left(t\right)$, which is defined by $\left\vert \left\langle \Psi \left( t\right)
|\psi \left( t\right) \right\rangle \right\vert ^{2}$, can be written as  
\begin{equation}
F\left( t\right) =\left\vert \sum\limits_{m=0}^{\infty }\left[ \cos \left( Jt\right)
A_{m}^{\ast }\left( t\right) -i\sin \left( Jt\right) B_{m}^{\ast }\left(
t\right) \right] \left\langle m\vert Z\left( t\right)
\right\rangle _{M}\right\vert ^{2}. \label{Ft}
\end{equation}

In Fig.~\ref{Figure2}(b), we plot the fidelity $F\left( t\right)$ as a function of time $t$ for different rescaled modulating amplitudes $\Delta_0/g_0$, under given detuning $\delta/g_0=1.13$.  Note that the squeezing of this given detuning corresponds to the red-solid curve in Fig.~\ref{Figure2}(a). We also would like to mention that the photon tunneling rate $J/g_0=398.6$ is chosen in order to get the strongest interference effect, which will be illustrated in Sec.~\ref{superpose}. Here, we focus on the evolution of $F\left( t\right)$ within $T_M=T/2=\pi/2\chi$, considering that it is long enough for us to get the macroscopically squeezed state with enhanced degree of squeezing. It can be seen that higher fidelity is arrived with the increase of the rescaled modulating amplitude $\Delta_0/g_0$, e.g., from 20 (blue line) ranges to 60 (red line), and then 100 (black line). However, it does not mean that the modulating amplitude $\Delta_0$ is allowed to increase unlimitedly, which actually has been implied in Eq.~(\ref{condition}). Figure~\ref{Figure3} helps us get more intuitional understanding about its restriction by $J$. Here we focus on the fidelity $F\left( T_M\right)$ at time $T_M$, where we get the maximum squeezing, as denoted by the red-dashed line in Fig.~\ref{Figure2}(a). In Fig.~\ref{Figure3}, we show the response of $F\left( T_M\right)$ to the variation of rescaled modulating amplitude $\Delta_0/g_0$, under different modulating frequencies $J/g_0$. Meanwhile, we fix the rescaled detuning $\delta/g_0=1.13$, and adjust the mechanical frequency following the relation $\omega_M=\delta+J$. It shows that high and stable fidelity $F\left( T_M\right)$ can only be obtained within a certain range of $\Delta_0/g_0$. Moreover, the range shrinks with the reduction of the rescaled tunneling rate $J/g_0$, e.g., from 398.6 (black solid line) to 258.6 (red dash-dotted line), and then 118.6 (blue dotted line). Once the rescaled modulating amplitude $\Delta_0/g_0$ goes out of the range, the fidelity $F\left( T_M\right)$ decreases in an oscillating way to zero, which means the validity of RWA needs to be reconsidered. Obviously, the characteristics shown in Fig.~\ref{Figure2}(b) and Fig.~\ref{Figure3} are in good accordance with the condition in Eq.~(\ref{condition}).

\begin{figure}[ptb]
\includegraphics[scale=0.39,clip]{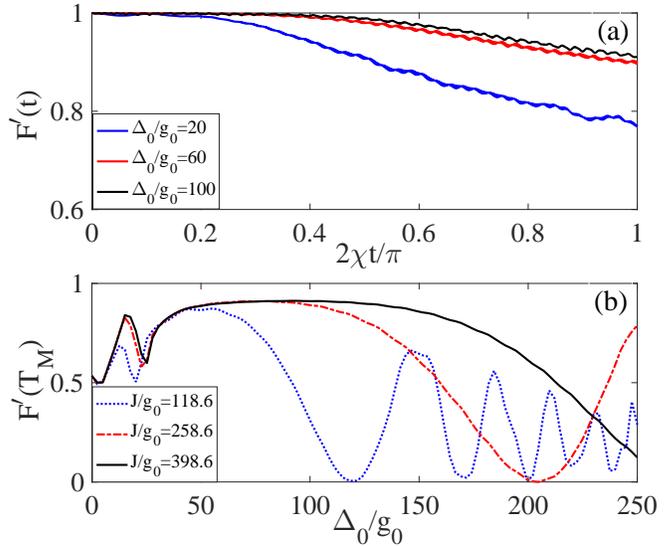}\
 \caption{(Color online) (a) The fidelity $F^{\prime}(t)$ for the given detuning $\delta/g_0=1.13$, under different rescaled modulating amplitude $\Delta_0/g_0=20,\,60,\,100$. The rescaled mechanical frequency $\omega_M/g_0=400$. (b) The fidelity $F^{\prime}(T_M)$ at the time $T_M$, where the maximum squeezing happens, versus the rescaled modulating amplitude $\Delta_0/g_0$, under different rescaled photon tunneling rates $J/g_0=118.6,\, 258.6, \,398.6$. Here the rescaled detuning $\delta/g_0$ is settled down to 1.13. The mechanical frequency is determined by $\omega_M=\delta+J$ for different tunnel rates $J$.}%
\label{Figure4}%
\end{figure}

The high fidelity under the condition in Eq.~(\ref{condition}), makes it reliable to generate the mechanical squeezed state with enhanced squeezing degree in our approach. Provided that the system was initialized as $\left\vert \psi \left( 0\right) \right\rangle=\left\vert 1\right\rangle_L\left\vert 0\right\rangle_R\left\vert 0\right\rangle_M$ and evolved for time $T_M$, the mechanical resonator can be projected into an enhanced squeezed vacuum state $\left\vert Z(T_M)\right\rangle$. Meanwhile, the degree of squeezing can be enhanced to 12.16 dB and even further. Next, we would like to show the superposition of the squeezed state with enhanced squeezing degree and its interference property.

\section{Generation of superposed enhanced squeezed states}\label{superpose}
In this section, we focus on the superposition of the macroscopically squeezed state with enhanced squeezing degree, which has been discussed in Sec.~\ref{enhance}. Similarly, we now assume that the initial state of the whole system has the form of 
\begin{equation}
\left\vert \psi^{\prime} \left( 0\right) \right\rangle=\frac{1}{\sqrt2}( \left\vert 1\right\rangle_L\left\vert 0\right\rangle_R+\left\vert 0\right\rangle_L\left\vert 1\right\rangle_R) \left\vert 0\right\rangle_M,
\end{equation}
which means that the photons in the left and right cavities are entangled with each other, and the mechanical resonator is initialized in the ground state. Utilizing the unitary evolution operator $U(t)$ in Eq.~(\ref{Ut}) and following the evolution equation in Eq.~(\ref{psitTheory}), the state $\left\vert \psi^{\prime} \left( t\right) \right\rangle$ of the system at time $t$ can be obtained analytically as 
\begin{eqnarray}
\left\vert \psi^{\prime} \left( t\right) \right\rangle&=&\frac{e^{-i\left[ \theta
\left( t\right) +\omega _{c}t\right] }}{\sqrt{2}}[ \left\vert
1\right\rangle _{L}\left\vert 0\right\rangle _{R}\left\vert \varphi
_{L}\left( t\right) \right\rangle _{M}\nonumber\\
&&+\left\vert 0\right\rangle
_{L}\left\vert 1\right\rangle _{R}\left\vert \varphi _{R}\left( t\right)
\right\rangle _{M}].\label{psipt}
\end{eqnarray}
Here, the global phase factor $\theta(t)$ has been defined in Eq.~(\ref{thetat}), and
\begin{eqnarray}
\left\vert \varphi _{L}\left( t\right) \right\rangle _{M} &=&\cos \left(
Jt\right) \left\vert Z\left( t\right) \right\rangle _{M}-ie^{i\Delta_{0}t}\sin \left(
Jt\right) \left\vert -Z\left( t\right) \right\rangle _{M},\nonumber \\
\left\vert \varphi _{R}\left( t\right) \right\rangle _{M} &=&e^{i\Delta_{0}t}\cos \left(
Jt\right) \left\vert -Z\left( t\right) \right\rangle _{M}-i\sin \left(
Jt\right) \left\vert Z\left( t\right) \right\rangle _{M}, \label{phiLR}\nonumber \\
\end{eqnarray}
represent two Yurke-Stoler-like~\cite{Yurke1986,Yurke1997,Yurke1990} quantum superpositions of the enhanced squeezing states with different relative phases, respectively. Following the same procedure in the last subsection, i.e., by measuring the photon in the left or right cavities at time $T_M$, we can get the mechanical resonator in the superposed squeezed states $\left\vert \varphi _{L}\left(T_M\right) \right\rangle _{M}$ or $\left\vert \varphi _{R}\left(T_M\right) \right\rangle _{M}$. 

Meanwhile, the numerical solution $\left\vert \Psi^{\prime} \left( t\right) \right\rangle$ can also be obtained from the corresponding initial condition $\left\vert \Psi^{\prime} \left( 0\right) \right\rangle=\left\vert \psi^{\prime} \left( 0\right) \right\rangle$, i.e. 
\begin{equation}
A_0(0)=B_0(0)=\frac{1}{\sqrt{2}}, \text{\ \ \ }A_{m>0}(0)=B_{m>0}=0,
\end{equation}
and the evolution equations shown in Eqs.~(\ref{Amt}) and~(\ref{Bmt}). In Figs.~\ref{Figure4}(a) and~\ref{Figure4}(b), the reliability of $\left\vert \psi^{\prime} \left( t\right) \right\rangle$ in Eq.~(\ref{psipt}) is verified by checking the fidelity dynamically and instantaneously, i.e., $F^{\prime}(t)$ and $F^{\prime}(T_M)$ under different conditions, respectively. Corresponding to the conditions in Eq.~(\ref{condition}), they behave similarly with $F(t)$ and $F(T_M)$ in Fig.~\ref{Figure2}(b) and Fig.~\ref{Figure3}, respectively, except for different absolute amplitudes and ranges. For example, under given photon tunneling rate $J$, the reliable range of modulating amplitudes $\Delta_0$ in Fig.~\ref{Figure4}(b) is smaller than that in Fig.~\ref{Figure3}, this may be induced by the interference between different channels shown in Eq.~(\ref{psipt}) and Eq.~(\ref{phiLR}).

\begin{figure}[ptb]
\includegraphics[scale=0.36,clip]{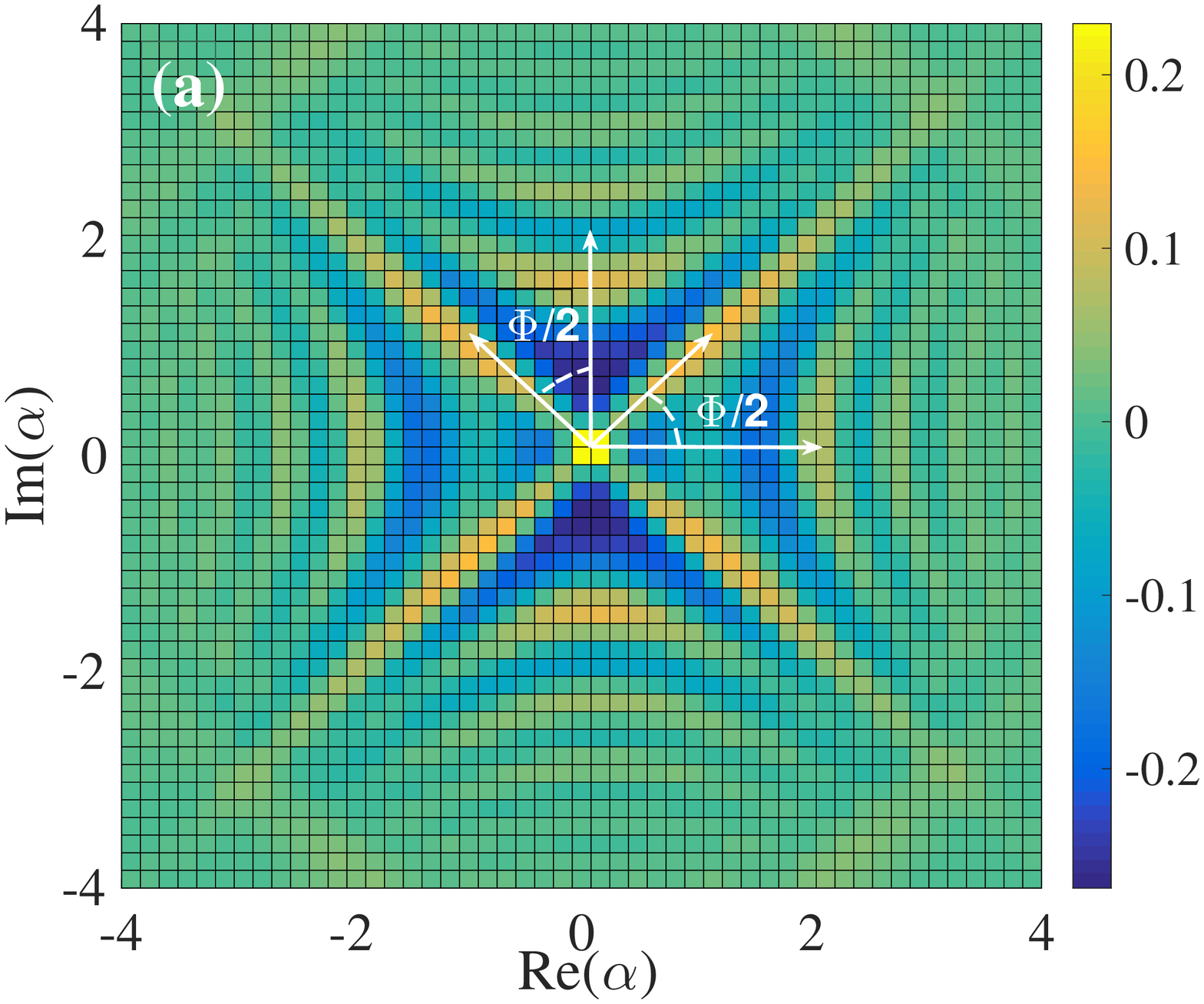}\
\includegraphics[scale=0.36,clip]{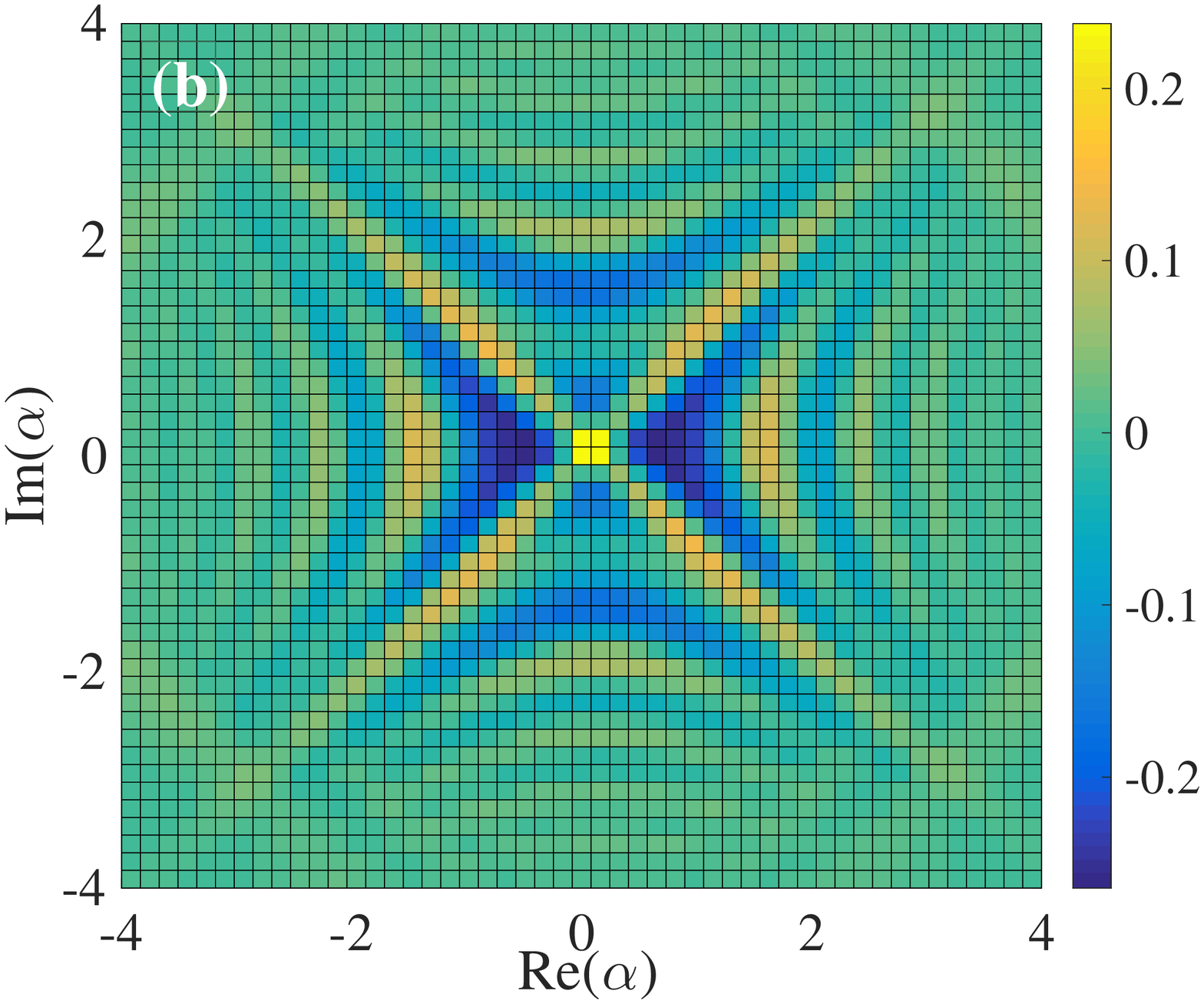}\
 \caption{(Color online) Wigner functions (a) $W_L(\alpha)$ and (b) $W_R(\alpha)$  for the superposed squeezed states $\left\vert \varphi _{L}\left( T_M\right) \right\rangle _{M}$ and $\left\vert \varphi _{R}\left( T_M\right) \right\rangle _{M}$, respectively. The other parameters are $J/g_0=398.6$, $\Delta_0/g_0=100$ and $\delta/g_0=1.13$. }%
\label{Figure5}%
\end{figure}

The interference effect in the superposed squeezed states can be seen from their Wigner quasi-probability distributions, which are defined as~\cite{Buzek1995,Barnett1997,Walls1995}
\begin{equation}
W_i(\alpha)=\frac{2}{\pi}{\rm{Tr}}[D^{\dagger}\left(\alpha\right)\rho_i D\left(\alpha\right)\left(-1\right)^{b^{\dagger}b}]. \label{Wi}
\end{equation}
Here, $D\left(\alpha\right)={\rm{exp}}\left(\alpha b^{\dagger}-\alpha^{*}b\right)$ is the displacement operator, and $\rho_i$ is the density operator of state $\left\vert \varphi_i \left(T_M\right) \right\rangle _{M}$ with the subscript $i=L, R$. Substituting Eq.~(\ref{phiLR}) into Eq.~(\ref{Wi}), the Wigner functions for the superposed squeezed states $\left\vert \varphi _{L}\left( T_M\right) \right\rangle _{M}$ and $\left\vert \varphi _{R}\left( T_M\right) \right\rangle _{M}$, can be written as
\begin{eqnarray}
W_{L}\left(\alpha\right) &=&\cos^{2}(JT_M) W_{+}+\sin^2(JT_M) W_{-}
\nonumber\\
&&+i\frac{\sin(2JT_M)}{2}(e^{-i\Delta_0T_M}W_{+-}-e^{i\Delta_0T_M}W_{-+}), \nonumber\\ 
W_{R}\left(\alpha\right) &=&\sin^{2}(JT_M) W_{+}+\cos^2(JT_M) W_{-}\nonumber
\end{eqnarray} 
\begin{equation}
+i\frac{\sin(2JT_M)}{2}(e^{i\Delta_0T_M}W_{-+}-e^{-i\Delta_0T_M}W_{+-}), \label{WLR}
\end{equation}
\begin{figure}[ptb]
\includegraphics[scale=0.38,clip]{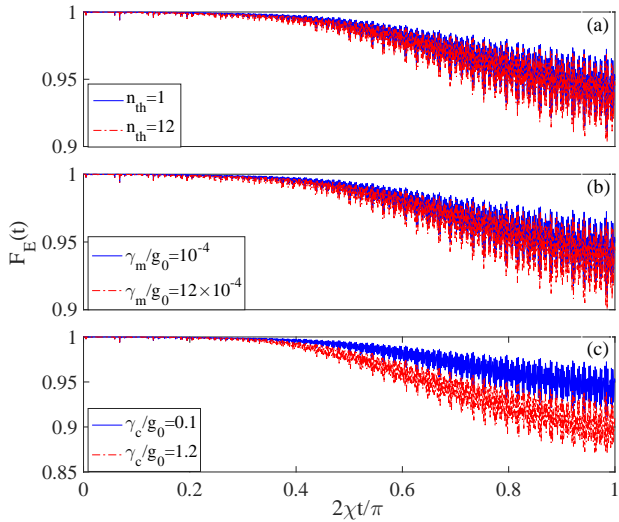}\
 \caption{(Color online) Fidelity $F_{\rm{E}}(t)$ versus time with different (a) thermal phonon numbers $n_{\rm{th}}$, (b) mechanical decay rates $\gamma_m$, and (c) cavity decay rates $\gamma_c$ in the open system, respectively. The reference case is $n_{\rm{th}}=1$, $\gamma_m/g_0=10^{-4}$, and $\gamma_c/g_0=0.1$, as denoted by the blue solid curves in (a), (b) and (c). When one of them is varied, the others remain unchanged. Other parameters are $J/g_0=398.6$, $\Delta_0/g_0=100$ and $\delta/g_0=1.13$. }%
\label{Figure6}%
\end{figure}
respectively. The parameter $W_{\pm}$ denotes the Wigner function of the squeezed state $\left \vert \pm Z(T_M)\right\rangle_M$, while $W_{\pm\mp}$ throws light on the quantum interference between two different squeezed states. For the compactness of the paper, the explicit expressions of $W_{+}$, $W_{-}$, $W_{+-}$ and $W_{-+}$ are shown in Appendix B. In Figs.~\ref{Figure5}(a) and~\ref{Figure5}(b), we show the Wigner functions $W_L(\alpha)$ and $W_R(\alpha)$ for the superposed squeezed states $\left\vert \varphi _{L}\left( T_M\right) \right\rangle _{M}$ and $\left\vert \varphi _{R}\left( T_M\right) \right\rangle _{M}$, respectively. In each figure, we find two error ellipses, which mainly come from the contribution of $W_{\pm}$, rotated from the horizontal axis by $\Phi/2$ and $(\Phi+\pi)/2$, respectively. The quantum fluctuation is strongly squeezed by $e^{-{R}}$ in one quadrature, while stretching by $e^{{R}}$ in the other one. Besides, the fringes appear around the two error ellipses, mainly come from the contribution of $W_{\pm\mp}$, denoting the quantum interference between the two different squeezed states. Moreover, it is obvious that the Wigner function $W_{R}\left(\alpha\right)$ in Fig.~\ref{Figure5}(b), can be obtained by a rotation of $\pi/2$ from $W_{L}\left(\alpha\right)$ in Fig.~\ref{Figure5}(a). This phenomenon actually is consistent with the relation, i.e., $\left\vert \varphi _{R}\left( T_M\right) \right\rangle _{M}=-e^{-ib^{\dagger}b\pi/2}\left\vert \varphi _{L}\left( T_M\right) \right\rangle _{M}$, which can be inferred from Eq.~(\ref{phiLR}). We would also like to mention that, without loss of generality, in Fig.~\ref{Figure5}, we consider the maximum interference by choosing the tunneling rate $J$ satisfying $\left\vert\cos\left(JT_M\right)/ ie^{i\Delta_{0}t}\sin \left(JT_M\right)\right\vert\simeq1$. Till now, we have shown delicately the generations and quasi-probability distributions of Yurke-Stoler-like~\cite{Yurke1986,Yurke1997,Yurke1990} enhanced squeezed vacuum states in the closed system. Actually, the method we used here can be extended to more general cases, e.g., generating the superposition of squeezed coherent or number states of mechanical resonator, but here we mainly focus on and take the vacuum case as an example. Next, we would like to study the effect of the environment fluctuations on the performance of this state generation scheme. 
\begin{figure}[ptb]
\includegraphics[scale=0.75,clip]{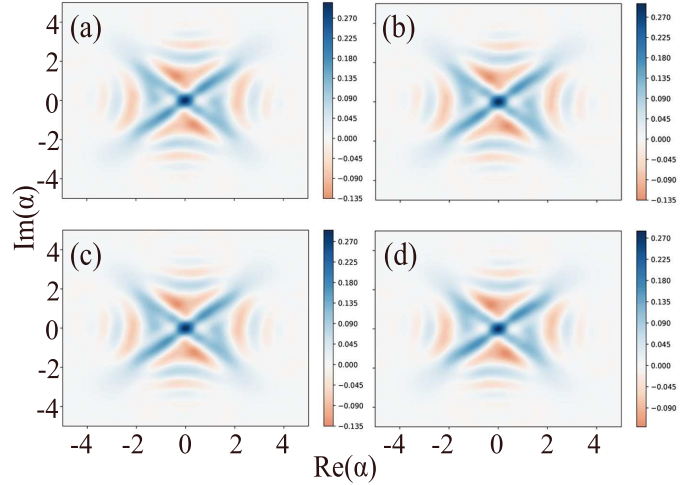}\
 \caption{(Color online) Wigner functions $W_{\rm{E}}(\alpha)$ for the superposed enhanced squeezed states $\left\vert \varphi _{L}\left( T_M\right) \right\rangle _{M}$ in the open system. The quasi-probability distribution of the reference case, which corresponds to $n_{\rm{th}}=1$, $\gamma_m/g_0=10^{-4}$, and $\gamma_c/g_0=0.1$, is shown in (a), while the cases when one of the dissipative parameters is varied is shown in (c-d). For example, the case corresponding to $n_{\rm{th}}=12$ is shown in (b), $\gamma_m/g_0=12\times10^{-4}$ in (c), and $\gamma_c/g_0=1.2$ in (d).} 
\label{Figure7}%
\end{figure}
\section{EFFECTS OF DISSIPATIONS} \label{dissipation}
Let us now study the effects of environmental fluctuations on the state generation, including the fidelity and Wigner quasi-probability distribution in open system. 

We study the environmental effect by using the master equation and including the dissipative terms, which can be expressed in the form of Lindblad superoperator, i.e., 
\begin{eqnarray}
\mathcal{L}_{\rm{diss}}&\simeq&\frac{\gamma_{m}}{2} \left[\left(n_{\rm{th}}+1\right)\mathcal{D}\left[b\right]+n_{\rm{th}}\mathcal{D}\left[b^{\dagger}\right] \right] \nonumber\\
  &&+\gamma_{L}\mathcal{D}\left[a_L\right]+\gamma_{R}\mathcal{D}\left[a_R\right]. \label{Ldiss}
\end{eqnarray}
Here, $\gamma_{L}$, $\gamma_{R}$, and $\gamma_{m}$ represent the decay rates of the left, right cavity modes and mechanical resonator, respectively. For simplicity and without loss of generality, we will assume that the photons in left and right cavities decay in the same rate in the following part of our paper, i.e., $\gamma_{L}=\gamma_{R}=\gamma_{c}$. The superoperator has the standard form of 
\begin{equation}
\mathcal{D}\left[o\right]\rho=o\rho o^{\dagger}-\left(o^{\dagger}o\rho+\rho o^{\dagger}o\right)/2,
\end{equation}
where $\rho$ is the reduced density matrix of the system, and $o$ can be any system operator, i.e., $a_L, a_R, b$.  The first line in Eq.~(\ref{Ldiss}) describes the coupling of the mechanical resonator to a thermal bath, and $n_{\rm{th}}=1/[\exp(\hbar\omega_{m}/k_{B}T)-1]$ denotes the thermal phonon number at temperature $T$, with $k_{B}$ the Boltzmann constant. The superoperators $\mathcal{D}\left[a_L\right]$ and $\mathcal{D}\left[a_R\right]$ represent the leakages of the cavity modes $a_L$ and $a_R$, respectively. Here, we assume that the cavity frequencies are much higher than the mechanical frequency, such that the thermal effect on the cavities can be neglected. Thus, the evolution of the system can be given by the quantum master equation, i.e.,
\begin{eqnarray}
\dot{\rho}=i\left[ \rho,H\left(t\right)\right]+\mathcal{L}_{\rm{diss}}\rho,\label{rho}
\end{eqnarray}
with the full Hamiltonian $H\left(t\right)$ given in Eq.~(\ref{H}). Once the initial condition corresponding to the initial state is given, e.g., $\rho\left(0\right)=\left\vert \psi^{\prime} \left( 0\right) \right\rangle\left\langle \psi^{\prime} \left( 0\right) \right\vert$, Eq.~(\ref{rho}) can be solved numerically in the complete basis set $\left\vert n_{L}\right\rangle\otimes \left\vert n_{R}\right\rangle\otimes \left\vert n_{b}\right\rangle$. Here, $n_{L}$, $n_{R}$, and $n_{b}=0,1,2\cdots$ denote the excitation numbers in cavity modes $a_L, a_R$, and mechanical resonator mode $b$, respectively. In this work, the numerical solution $\rho\left(t\right)$ of the master equation in Eq.~(\ref{rho}) can be obtained utilizing the quantum toolbox~\cite{Johansson2012,Johansson2013} within a truncated Fock state space.

In Fig.~\ref{Figure6} and Fig.~\ref{Figure7}, we show the effects of environmental fluctuations on the fidelity $F_{\rm{E}}(t)$ and Wigner distribution $W_{\rm{E}}(\alpha)$, respectively. The subscript E denotes that the effect of the environment is involved. For the purpose of convenience, here we only take the state $\left\vert \varphi _{L}\left( T_M\right) \right\rangle _{M}$ as an example, and the environment has similar effect on the state $\left\vert \varphi _{R}\left( T_M\right) \right\rangle _{M}$. In Figs.~\ref{Figure6}(a-c) and Figs.~\ref{Figure7}(a-d), we consider different thermal phonon numbers $n_{\rm{th}}$, mechanical decay rates $\gamma_m$ and cavity decay rates $\gamma_c$, respectively. We take the case $n_{\rm{th}}=1$, $\gamma_m/g_0=10^{-4}$, and $\gamma_c/g_0=0.1$ as a reference, whose fidelity and Wigner function are denoted by the blue solid lines in Figs.~\ref{Figure6}(a-c) and the quasi-probability distribution in Fig.~\ref{Figure7}(a), respectively. Meanwhile, the cases when one of the dissipative parameters is varied, are denoted by the red dash-dotted lines in Figs.~\ref{Figure6}(a-c) and the distributions in Figs.~\ref{Figure7}(b-d). For example, the fidelity and Wigner function corresponding to $n_{\rm{th}}=12$ are shown in Figs.~\ref{Figure6}(a) and~\ref{Figure7}(b), $\gamma_m/g_0=12\times10^{-4}$ in Figs.~\ref{Figure6}(b) and~\ref{Figure7}(c), and $\gamma_c/g_0=1.2$ in Figs.~\ref{Figure6}(c) and~\ref{Figure7}(d).


Comparing the unperturbed cases in Figs.~\ref{Figure4}(a) and~\ref{Figure5}(a) with the perturbed ones in Figs.~\ref{Figure6} and~\ref{Figure7} respectively, we can see great correspondence in the overall behavior, except for larger and faster oscillation, which are induced by faster photon and phonon leakages. Besides, in Figs.~\ref{Figure6}(a) and~\ref{Figure6}(b), when $n_{\rm{th}}$ and $10^{4}\gamma_M/g_0$ are increased from $1$ (blue-solid curve), to $12$ (red dash-dotted curve), the fidelity $F_{\rm{E}}(t)$ remains almost unchanged. The corresponding Wigner distributions in Figs.~\ref{Figure7}(a), (b), (c) show almost the same patterns and absolute probability amplitudes. The system shows slightly obvious but reliability guaranteed degradation due to the increase of cavity dissipation, as shown in the decreasing amplitudes of the red dash-dotted curve in Fig.~\ref{Figure6}(c) and quasi-probabilities in Fig.~\ref{Figure7}(d). This phenomenon can be attributed to the fact that the photon leakage is the foremost resource of system dissipation. 

Moreover, our approach remains robust even with further strong dissipation, e.g., $n_{\rm{th}}=50$, $\gamma_m/g_0=50\times10^{-4}$, or $\gamma_c/g_0=5$, which are checked but not included in our paper for the clearness of figures. When the system dissipations are further increased, the quasi-probability amplitudes show small change, e.g., the amplitudes get smaller in diagonal areas while larger in non-diagonal interference fringes, which means the probability distribution of the system becomes diffused. However, the overall patterns remain similar with less disturbed cases. Thus, we can see that this state generation scheme shows robust performance and has good resistance to environmental dissipations.

\section{CONCLUSIONS}\label{conclusion}
In summary, we study the superposition of the macroscopically squeezed states with enhanced squeezing in a weakly coupled two-mode optomechanical system. Notably enhanced squeezing (12.16dB) for the mechanical resonator can be achieved with single-photon, in the way of modulating either the cavity frequencies or the photon hopping rate and optomechanical coupling strength. In this work, we mainly focus on the first case, i.e., the system with modulated cavity frequencies.

Once the modulating amplitude and frequency satisfy certain conditions, it is reasonable to conduct rotating wave approximation (RWA) and neglect far-off-resonant terms. Meanwhile, the system with near-resonant terms can be approximated by an effective quadratically coupled two-mode optomechanical system with high fidelity. The Hamiltonian with flexibly modified ratio of coupling strength indicates that further considerable larger squeezing can be obtained by carefully adjusting the system parameters.  

Moreover, by initializing the two cavity modes into proper states and measuring the state of photon after time $T_M$, the mechanical resonator can be designed into a superposition state. In terms of Wigner quasi-probability distribution, we show the enhanced squeezing and superposition properties of the Yurke-Stoler-type~\cite{Yurke1986,Yurke1997,Yurke1990} squeezed state, from the squeezed error ellipses and interference fringes, respectively. We also verify the validity of our approach in both the closed and open systems, i.e., without and with the dissipative fluctuations. The fidelity and Wigner distribution remain almost unchanged with small perturbations, while show little diffusion but with reliability guaranteed and pattern protected under further strong dissipation. 

Our research show that this state generation scheme is reliable and exhibit comparable tolerance to environmental fluctuations. The generated states, which have enhanced squeezing and superposition properties, hold great promise for both fundamental interest, e.g., quantum superposition principle, and practical value, e.g., ultrasensitive measurement.   

\begin{acknowledgments}
Y.X.L. acknowledges the financial support of the National Basic Research Program of China 973 Program under Grant No.~2014CB921401, the Tsinghua University Initiative Scientific Research Program, and the Tsinghua National Laboratory for Information Science and Technology (TNList) Cross-discipline Foundation. Y. Zhang would like to acknowledge the financial support of the NSF of China under Grants No.~11674201.
\end{acknowledgments}

\appendix
\section{MODULATED COUPLING STRENGTHS}\label{Mcoupling}
In this appendix, we consider the case when sinusoidal modulations are applied both to the photon-hopping interaction $J$ and quadratically coupling strength $g_0$, but with different modulating strength and frequency. Then the Hamiltonian has the form of 
 \begin{eqnarray}
H^{\prime}\left( t\right) &=&\omega _{c}\left( a_{L}^{\dagger }a_{L}+a_{R}^{\dagger
}a_{R}\right)+J(t) \left(
a_{L}^{\dagger }a_{R}+a_{R}^{\dagger }a_{L}\right) \nonumber\\
&&+\omega _{M}b^{\dagger }b+g(t) \left( a_{L}^{\dagger
}a_{L}-a_{R}^{\dagger }a_{R}\right) \left( b^{\dagger }+b\right) ^{2}, \nonumber\\
\end{eqnarray}
where the time-dependent coupling strengths
\begin{equation}
J(t)=J\omega _{0}\cos \left( \omega _{0}t\right),\ \ \ g(t)=g_{0}\cos \left( 2\omega _{p}t\right).
\end{equation}
Similarly, to see clearly the effect of coupling strength modulation, we perform a series of transformation defined by
\begin{eqnarray}
V_{1}\left( t\right) &=&e^{-i\left[ \omega _{c}\left( a_{L}^{\dagger
}a_{L}+a_{R}^{\dagger }a_{R}\right) +\omega _{M}b^{\dagger }b\right] t}, \nonumber\\
V_{2}\left( t\right) &=&e^{-iJ\sin \left( \omega _{0}t\right) \left(
a_{L}^{\dagger }a_{R}+a_{R}^{\dagger }a_{L}\right) }.
\end{eqnarray}
Then the transformed Hamiltonian can be written as
\begin{eqnarray}
H^{\prime}_{2}\left( t\right)&=&\left[ 
\begin{array}{c}
\cos \left( 2J \sin \left( \omega _{0}t\right) \right) \left(
a_{L}^{\dagger }a_{L}-a_{R}^{\dagger }a_{R}\right)  \\ 
+i\sin \left( 2J \sin \left( \omega _{0}t\right) \right) \left(
a_{R}^{\dagger }a_{L}-a_{R}a_{L}^{\dagger }\right) 
\end{array}%
\right]\nonumber\\ &&\times \frac{g_{0}}{2}\left[G_1(b)+\rm{H.c.}\right]. \label{H2pt}
\end{eqnarray}
Utilizing the Jacobi-Anger expansions 
\begin{eqnarray*}
\cos \left( 2J \sin \left( \omega _{0}t\right) \right)  &=&J_{0}\left(
2J \right) +2\sum\limits_{n=0}^{\infty }J_{2n}\left( 2J \right) \cos
\left( 2n\omega _{0}t\right), \\
\sin \left( 2J \sin \left( \omega _{0}t\right) \right) 
&=&2\sum\limits_{n=0}^{\infty }J_{2n-1}\left( 2J \right) \sin \left(
\left( 2n-1\right) \omega _{0}t\right), 
\end{eqnarray*}
with $J_m(x)$ being the Bessel function of the first kind and $m$ being an integer, the Hamiltonian $H^{\prime}_{2}(t)$ in Eq.~(\ref{H2pt}) can be expanded into
\begin{eqnarray}
H^{\prime}_{2}(t)&=&\frac{g_{0}}{2}J_{0}\left( 2\xi \right) \left( a_{L}^{\dagger
}a_{L}-a_{R}^{\dagger }a_{R}\right) \left[G_1(b)+\rm{H.c.}\right]  \nonumber\\
&&+\frac{g_{0}}{2}\left( a_{L}^{\dagger }a_{L}-a_{R}^{\dagger }a_{R}\right)
\sum\limits_{n=0}^{\infty }J_{2n}\left( 2\xi \right) \left[G_{2n}(b)+\rm{H.c.}\right]  \nonumber \\
&&+\frac{g_{0}}{2}\left( a_{R}^{\dagger }a_{L}-a_{R}a_{L}^{\dagger }\right)
\sum\limits_{n=0}^{\infty }J_{2n-1}\left( 2\xi \right)\left[G_{3n}(b)-\rm{H.c.}\right]. \nonumber\\
\end{eqnarray}
Here the functions have the form of
\begin{eqnarray}
G_1(b)&=&b^{2}\left(e^{-2i\delta_{-}^{\prime} t}+e^{-2i\delta_{+}^{\prime} t}\right)+\left( 2b^{\dagger }b+1\right)e^{2i\omega _{p}t},\\ 
G_{2n}(b)&=&b^{2}\left(e^{-2i\delta _{n--} t}+e^{-2i\delta _{n-+} t}+e^{-2i\delta _{n+-} t}+e^{-2i\delta _{n++} t}\right)\nonumber\\
&&+\left( 2b^{\dagger }b+1\right)\left(e^{2i\delta p_{n-}t}+e^{2i\delta p_{n+}t}\right),\\
G_{3n}(b)&=&b^{2}\left(e^{-2i\delta _{n--} t}+e^{-2i\delta _{n+-} t}\right)e^{-i\omega_0t}\nonumber\\
&&-\left(e^{-2i\delta _{n++} t}+e^{-2i\delta _{n-+} t}\right)e^{i\omega_0t}\nonumber\\
&&+\left( 2b^{\dagger }b+1\right)\left(e^{2i\delta p_{n+}t}e^{-i\omega_0t}-e^{2i\delta p_{n-}t}e^{i\omega_0t}\right),
\end{eqnarray}
with the detunings defined as
\begin{eqnarray}
\delta^{\prime}_{\pm}&=&\omega_M\pm\omega_p,\\
\delta _{n\pm\pm}&=&\delta^{\prime}_{\pm}\pm n\omega_0,\\
\delta p_{n\pm}&=&\omega_p\pm n\omega_0.
\end{eqnarray}
Obviously, more frequency terms appear because of the modulation of coupling strength, but we can always identify fast oscillating terms under specific conditions and perform the rotating-wave approximation (RWA). Consider the case 
\begin{equation}
\delta^{\prime}, g^{\prime} \ll \omega_0, \omega_M-\omega_p, \omega_p, \label{condition2}
\end{equation}
with 
\begin{equation}
\delta^{\prime}=\delta _{n_0--}, \ \ \  g^{\prime}=\frac{g_0}{2}J_{2n_0}(2J), 
\end{equation}
then we can arrive at 
\begin{equation}
\tilde{H}_{2}\left( t\right) \simeq g^{\prime}\left( a_{L}^{\dagger
}a_{L}-a_{R}^{\dagger }a_{R}\right) \left( b^{2}e^{-2i\delta^{\prime} t}+b^{\dagger
2}e^{2i\delta^{\prime} t}\right), \label{Heff2}
\end{equation}
which is in the same form with Eq.~(\ref{Heff}).

 \section{INTERFERENCE EFFECT IN WIGNER DISTRIBUTION}\label{wigner}
 In this section, we show the details in the process of getting Eq.~(\ref{WLR}). Utilizing Eq.~(\ref{phiLR}), Eq.~(\ref{Wi}) can be written in the form of Eq.~(\ref{WLR}), where the explicit expressions of $W_{+}$, $W_{-}$, $W_{+-}$ and $W_{-+}$ can be written deliberately as
 \begin{eqnarray}
W_{+}&=&\frac{2}{\pi}\rm{Tr}[D^{\dagger}\left(\alpha\right)\left\vert Z\right\rangle \left\langle Z\right\vert D\left(\alpha\right)\left(-1\right)^{b^{\dagger}b}]\nonumber\\
&=&\frac{2}{\pi}\sum_{k}\left(-1\right)^k\left\vert S\left(k,-\alpha,Z\right) \right\vert^2;\\
W_{+-}&=&\frac{2}{\pi}\rm{Tr}[D^{\dagger}\left(\alpha\right)\left\vert Z\right\rangle \left\langle -Z\right\vert D\left(\alpha\right)\left(-1\right)^{b^{\dagger}b}]\nonumber\\
&=&\frac{2}{\pi}\sum_{k}\left(-1\right)^k S\left(k,-\alpha,Z\right)S^{*}\left(k,-\alpha,-Z\right);\\
W_{-+}&=&\frac{2}{\pi}\rm{Tr}[D^{\dagger}\left(\alpha\right)\left\vert -Z\right\rangle \left\langle Z\right\vert D\left(\alpha\right)\left(-1\right)^{b^{\dagger}b}]\nonumber\\
&=&\frac{2}{\pi}\sum_{k}\left(-1\right)^k S\left(k,-\alpha,-Z\right)S^{*}\left(k,-\alpha,Z\right);\\
W_{-}&=&\frac{2}{\pi}\rm{Tr}[D^{\dagger}\left(\alpha\right)\left\vert -Z\right\rangle \left\langle -Z\right\vert D\left(\alpha\right)\left(-1\right)^{b^{\dagger}b}]\nonumber\\
&=&\frac{2}{\pi}\sum_{k}\left(-1\right)^k\left\vert S\left(k,-\alpha,-Z\right) \right\vert^2; 
 \end{eqnarray}
 Here, the parameter $S\left(k,\alpha,Z\right)$ denotes the amplitude of finding $k$ phonons in the squeezed coherent state $\left\vert \alpha, Z\right\rangle$~\cite{Gerry2004}, i.e., 
\begin{eqnarray}
S\left(k,\alpha,Z\right)&=&\left\langle k \vert \alpha, Z\right\rangle \nonumber\\
&=&\frac{1}{\sqrt{\cosh R}}\rm{exp}\left[-\frac{1}{2}\left\vert\alpha \right\vert^2-\frac{1}{2}\alpha^{*2}e^{i\theta}\tanh R\right]\nonumber\\
&&\times\frac{\left[ \frac{1}{2}e^{i\theta}\tanh R\right]^{k/2}}{\sqrt{n!}}H_n\left[\gamma \left( e^{i\theta}\sinh(2R)\right)^{-1/2} \right] \nonumber\\
\end{eqnarray}

\end{document}